\documentclass[journal]{IEEEtran}
\ifCLASSINFOpdf
  \usepackage[pdftex]{graphicx}
\else
\fi
\usepackage{amsmath}
\usepackage{lipsum}

\usepackage{graphicx}

\usepackage{url}

\begin{document}
%
\title{Lung Sound Classification Using Co-tuning and Stochastic Normalization}
%
%
%

\author{T. Nguyen, and F. Pernkopf, \IEEEmembership{Member, IEEE}
\thanks{Thanks to the Vietnamese - Austrian Government scholarship and the Austrian Science Fund (FWF) under the 
	project number I2706-N31.}
\thanks{T. Nguyen, is with Signal Processing and Speech Communication Lab.
	Graz University of Technology, Austria (e-mail: t.k.nguyen@tugraz.at).}
\thanks{F. Pernkopf is with Signal Processing and Speech Communication Lab.
Graz University of Technology, Austria (e-mail: pernkopf@tugraz.at).}}

\maketitle

\begin{abstract}
In this paper, we use pre-trained ResNet models as backbone architectures for classification of adventitious lung sounds and respiratory diseases. The knowledge of the pre-trained model is transferred by using vanilla fine-tuning, co-tuning, stochastic normalization and the combination of the co-tuning and stochastic normalization techniques. Furthermore, data augmentation in both time domain and  time-frequency  domain is used to account for the class imbalance of the ICBHI and our multi-channel lung sound dataset. Additionally, we apply spectrum correction to consider the variations of the recording device properties on the ICBHI dataset. Empirically, our proposed systems mostly outperform all state-of-the-art lung sound classification systems for the adventitious lung sounds and respiratory diseases of both datasets. 
\end{abstract}


%
\IEEEpeerreviewmaketitle

\section{Introduction}
%
%
%
%
\IEEEPARstart{R}{espiratory} diseases have become one of the main causes of death in  society.  According to the World Health Organization (WHO), the "big five"
respiratory diseases, which include asthma, chronic obstructive pulmonary disease (COPD), acute lower respiratory tract infections, lung cancer and tuberculosis, cause the mortality of more than 3 million people each year worldwide. Currently, CoViD-19, a special form of viral pneumonia related to the coronavirus identified firstly in Wuhan (China) in 2019 \cite{chen2020epidemiological}, has caused globally more than 158 million infections and 3,296,000 deaths~\cite{whocovid19}. On March 11, 2020, the WHO officially announced that CoViD-19 has reached global pandemic status.  Furthermore, according to \cite{barbosa2021big}, the ‘‘big five’’ lung diseases, except lung cancer, have increased during CoViD-19 epidemics. These respiratory diseases are characterised by highly similar symptoms, i.e. the adventitious breathing, which could be a confounding factor during diagnosis \cite{chen2020epidemiological}. Due to their severe consequences, particularly in the case of CoViD-19, an early and accurate diagnosis of these types of diseases has become crucial. 

Lung sounds convey relevant information related to pulmonary disorders with adventitious breathing sounds such as crackles, wheezes, or both of crackles and wheezes~\cite{sarkar2015auscultation},~\cite{rocha2018alpha}. In the last decades, to facilitate a more objective assessment of the lung sound for diagnosis of pulmonary diseases/conditions,  computational methods i.e. computational lung sound analysis (CLSA)~\cite{gurung2011computerized},
~\cite{pramono2017automatic} have been developed. The CLSA systems automatically detect and classify adventitious lung sounds by using digital recording devices, signal processing techniques and machine learning algorithms. They are also carefully evaluated in real-life scenarios and can be used as portable easy-to-use devices without the necessity of expert interaction; especially, beneficial when facing infectious diseases as CoViD-19. 

In CLSA systems, there are two popular classification tasks, namely (i) adventitious lung sound and (ii) respiratory disease classification. In adventitious lung sound classification, recognition of normal and abnormal sounds (i.e. either crackles or wheezes or both of them) is important; while for respiratory disease classification, several categories  have been considered e.g. binary classification (health and pathological), ternary chronic classification (healthy, chronic and non-chronic diseases) or six class classification of distinct pathologies. The systems have been evaluated on non-public
datasets such as R.A.L.E.~\cite{rale1990} or multi channel lung sound data~\cite{messner2016robust} (ours) and public datasets i.e. the ICBHI 2017 dataset~\cite{rocha2018alpha} or the Abdullah University Hospital 2020 dataset~\cite{fraiwan2021dataset}. Due to limitations in the amount and quality of available data, the performance and generalization of the lung sound classification system may suffer over-estimated results. To deal with these challenges, different feature extraction methods~\cite{chambres2018automatic},~\cite{jakovljevic2017hidden},~\cite{chen2019triple},~\cite{pham2021cnn}, conventional machine learning~\cite{jakovljevic2017hidden},~\cite{malmberg1996SOM},~\cite{bahoura2009patterngmm},~\cite{bokov2016wheezingSVM},~\cite{liu2017lung}, deep learning~\cite{messner2018crackle},~\cite{dPerna2019lstm},~\cite{aykanat2017cnn},
~\cite{nguyen2020lung}, data augmentation and transfer learning from ImageNet~\cite{demir2020convolutional},~\cite{acharya2020deep}, or audio scene datasets~\cite{shi2019lung} have been explored. 

In this work, we improve the generalization ability and model performance for adventitious lung sound classification and respiratory disease classification systems using the ICBHI 2017 dataset and our multi-channel lung sound dataset. We exploit transfer learning approaches  such as co-tuning~\cite{you2020co} for different architectures of residual neural networks (ResNets). We use pre-trained ResNet models of the ImageNet classification task as backbone architectures, which requires a 3-channel input i.e. color RGB images. Therefore the spectrograms are converted into 3 channels for the model input. Particularly, logmel spectrograms are replicated into three channels for the adventitious lung sound task or converted into RGB color spectrogram for respiratory disease classification. The pre-trained models are exploited systematically in the following compositions. 
\begin{itemize}
	\item Firstly, we fine-tune the pre-trained model on a target domain and update all top (i.e. feature representation) layers and bottom (i.e. task-specific) layers. We call this vanilla fine-tuning.
	\item Secondly, we apply co-tuning for transfer learning~\cite{you2020co}, in which representation layers and task-specific layers of both source domain and target domain are collaboratively fine-tuned. Co-tuning further updates task-specific layers of the pre-trained model using a learned category relationship between source and target domains.
	\item Thirdly, we replace Batch Normalization (BN) layers, which suffer from poor performance in case of a data distribution shift between training and test data. We introduce stochastic normalization (StochNorm)~\cite{kou2020stochastic} in each residual block of the pre-trained backbone architecture. StochNorm is a parallel structure normalizing the activation of each channel by either mini-batch statistics or moving statistics to avoid influence of sample statistics during training. Thus, it is considered as a regularization method. Furthermore, fine-tuning inherits further prior knowledge of moving statistics of the pre-trained networks compared to vanilla fine-tuning. Both properties help to avoid over-fitting on small datasets such as the ICBHI and our lung sound dataset.
	\item Finally, we combine co-tuning and stochastic normalization techniques to take advantages of both techniques.
\end{itemize}
In addition,  we apply data augmentation in both time domain and time-frequency domain to account for the class imbalance in the datasets. Furthermore, we use spectrum correction for lung sound classification to compensate the recording device variations in the ICBHI dataset. The main contributions of the paper are:
\begin{itemize}
    \item We propose robust classification systems for adventitious lung sounds and respiratory diseases for the ICBHI and our multi-channel lung sound dataset.
	\item We exploit transferred knowledge of pre-trained models by vanilla fine-tuning, co-tuning, stochastic normalization techniques and a combination of both co-tuning and stochastic normalization.
	\item We introduce spectrum correction to improve the generalization ability by accounting for the recording device differences.
	\item In addition to commonly used data augmentation techniques, we double the size of the training dataset by flipping samples in target domain. This enhances the performance of adventitious lung sound classification.
	\item We review state-of-the-art adventitious lung sound and respiratory disease classification systems for the ICBHI and our multi-channel lung sound dataset.
\end{itemize}

The outline of the paper is as follows: In Section II, we introduce the lung sound databases. In Section III, we present our lung sound classification systems. In Section IV, we present the experimental setup including the evaluation metrics and the experimental results. We review related works in Section V. Finally, we conclude the paper in Section VI.

\section{Databases}
\subsection{ICBHI 2017 Dataset}

The ICBHI 2017 database~\cite{rocha2018alpha} consists of 920 annotated audio samples from 126 subjects corresponding to patient pathological conditions i.e. healthy and seven distinct disease categories (Pneumonia, Bronchiectasis, COPD, upper respiratory tract infection (URTI), lower respiratory tract infection (LRTI), Bronchiolitis, Asthma). The audios were recorded using different stethoscopes i.e. AKGC417L, Meditron, Litt3200 and LittC2SE. The recording duration ranges from 10s to 90s and the sampling rate ranges from 4000Hz to 44100Hz. Each recording is composed of a certain number of breathing cycles with corresponding annotations of the beginning and the end, and the presence/absence of crackles and/or wheezes. The annotations of the database supports to split audio recordings into respiratory cycles. The cycle duration ranges from 0.2s to 16s and the average cycle duration is 2.7s. The database includes 6898 different respiratory cycles with 3642 normal cycles, 1864 crackles, 886 wheezes, and 506 cycles containing of both crackles and wheezes.

We propose a classification system for the following tasks.
\begin{itemize}
   	\item ALSC: Adventitious lung sound classification (ALSC) is separated into two sub-tasks for respiratory cycles. The first one is a 4-class task classifying respiratory cycles into four classes (\textit{Normal}, \textit{Crackles}, \textit{Wheezes} and \textit{both Crackles and Wheezes}). The second sub-tasks is a 2-class task of normal and abnormal lung sounds including \textit{Crackles}, \textit{Wheezes} and \textit{both Crackles and Wheezes}. We evaluate our system on the official ICBHI data split. The dataset was divided by the ICBHI challenge into 60\% for training and 40\% for testing. Both sets are composed with different patients.
	 \item RDC: Respiratory disease classification (RDC) also consists of two sub-tasks for audio  recordings. The first one is a 3-class task classifying audio recordings into three groups of \textit{Healthy}, \textit{Chronic Diseases} (i.e. COPD, Bronchiectasis and Asthma) and \textit{Non-Chronic Diseases} (i.e. URTI, LRTI, Pneumonia and Bronchiolitis). The second sub-tasks is a 2-class task (healthy/unhealthy), where the unhealthy class comprises of the seven diseases.
\end{itemize}

\subsection{Multi-channel Lung Sound Database}
\begin{figure}[t]
\centering
\includegraphics[width=0.3\textwidth]{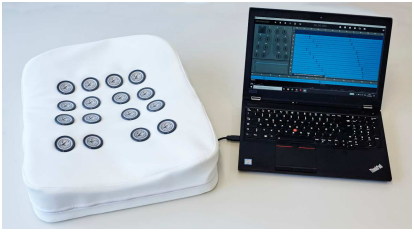}
\caption{Multi-channel lung sound recording device.}
\label{fig:recordingdevice}
\end{figure}

The multi-channel lung sound database~\cite{messner2016robust}, \cite{messner2018crackle}, \cite{messner2020multi} has been recorded in a clinical trial. It contains lung sounds of 16 healthy subjects and 7 patients diagnosed with idiopathic pulmonary fibrosis (IPF). We used our 16-channel lung sound recording device (see Fig.~\ref{fig:recordingdevice}) to record lung sounds over the posterior chest at two different airflow rates, with 3 - 8 respiratory cycles within 30s. The lung sounds were recorded with a sampling frequency of 16kHz. The sensor signals are filtered with a Bessel high-pass filter with a cut-off frequency of 80Hz and a slope of 24dB/oct. We extracted full respiratory cycles using the airflow signal from all recordings. We manually annotated respiratory cycles in cooperation with respiratory experts from Medical University of Graz, Austria. The number of breathing cycles with/without IPF are shown in Table~\ref{number_cycles}.

\begin{table}[h]
    \renewcommand{\arraystretch}{1.3}
    \caption{Number of subjects and cycles in the dataset}
    \label{number_cycles}
    \centering
    \scalebox{0.8}{
    \begin{tabular}{p{0.2\linewidth}p{0.1\linewidth}p{0.15\linewidth}p{0.15\linewidth}p{0.1\linewidth}} 
        \hline
        \multicolumn{2}{c}{\textbf{\# Subjects}}  & \multicolumn{3}{c}{\textbf{\# Respiratory Cycles}}\\
        \hline
        \textbf{Healthy} & \textbf{IPF} & \textbf{Normal} & \textbf{Crackles} & \textbf{Total}\\
        
        \hline
        16 & 7 & 4405 & 1791 & 6196 \\
        \hline
    \end{tabular}
}
\end{table}

\section{Transfer Learning for Lung sound Classification}
\begin{figure*}[t]
	\centering
    \includegraphics[width=\textwidth]{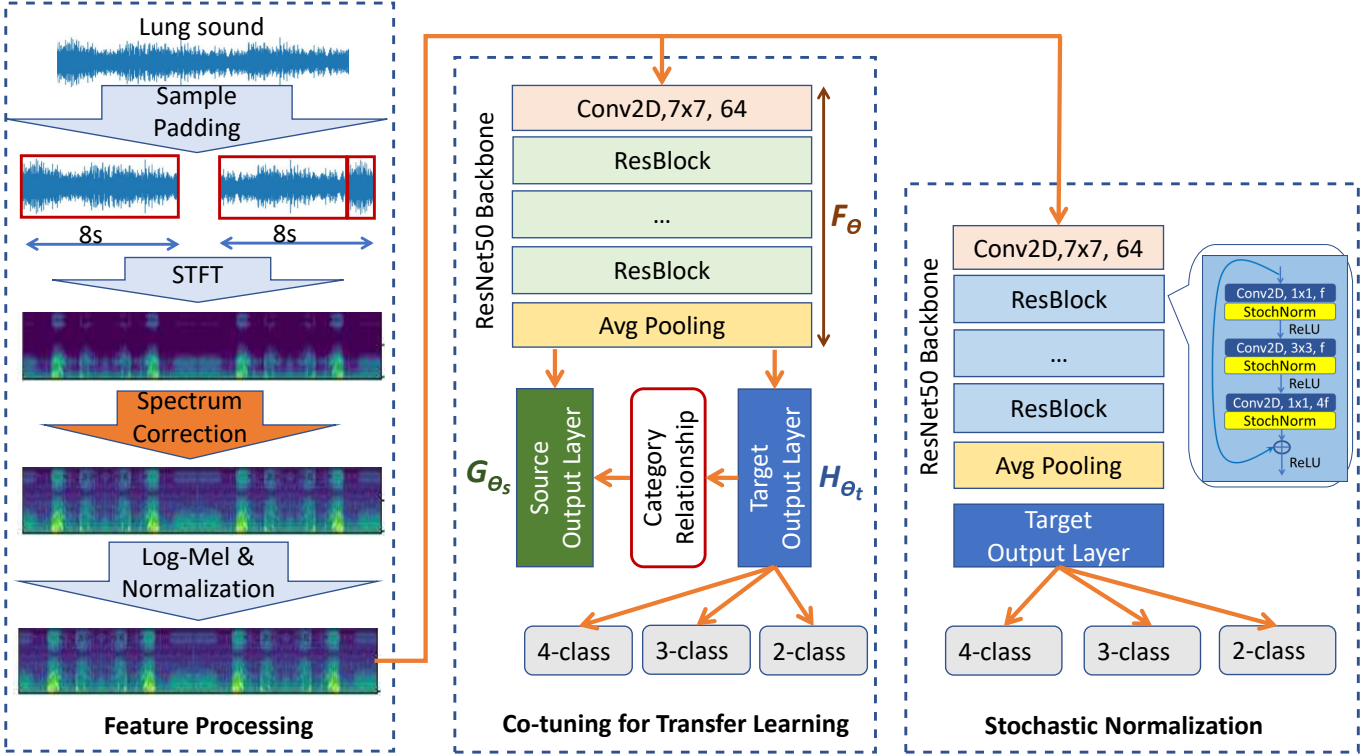}
	\caption{Proposed transferred knowledge systems using co-tuning for transfer learning or stochastic normalization.}
	\label{fig:framework}
\end{figure*}

The proposed systems include two key stages i.e. feature processing and classification as shown in Fig.~\ref{fig:framework}. Firstly, the respiratory cycles/ recordings are pre-processed in time domain and transformed into log-mel spectrograms of fixed size. Secondly, the features are fed to the CNN model where co-tuning or stochastic normalization are explored for the different classification tasks. During inference, the label of an input respiratory cycle/ recording is determined via majority voting~\cite{feldman1989majority} of the predicted labels of the individual segments with a length of 8 seconds belonging to the same recording. 

\subsection{Audio Pre-processing and Feature Extraction}
We use the audio pre-processing and feature extraction techniques presented in~\cite{nguyen2020lung} for both datasets. Audio recordings are resampled to 16kHz for the ALSC tasks of the ICBHI challenge and our dataset, while the RDC tasks use 4kHz of sampling rate. Similar to our previous works on ALSC of ICBHI and our multi-channel dataset~\cite{nguyen2020lung},~\cite{nguyen2021crackle}, the respiratory cycles are split without overlap into segments. Furthermore, we apply sample padding in time-reversed order to achieve fixed-length segments without abrupt signal changes. For the RDC task of the ICBHI dataset, recordings are split into segments of the same length using 50\% overlap. Different segment lengths are investigated in Section IV. Then we also applied sample padding to the segments being shorter than the fixed length. This means that for both respiratory cycles and recordings, the same splitting and sample padding procedure is applied to obtain the fixed-length segments for the ALSC and RDC tasks, respectively.

We use a window size of 512 samples for the fast Fourier transform (FFT) using 50\% overlap between the windows. The number of mel frequency bins is chosen as 50 and 45 for the ICBHI dataset and our multi-channel dataset, respectively. The logarithmic scale is applied to the magnitude of the mel spectrograms. The log-mel spectrograms are normalized with zero mean and unit variance. Then these spectrograms are duplicated into three channels to match the input size of the pre-trained ResNet model for the ALSC task. However, for the RDC task of the ICBHI dataset, we convert the spectrogram into a RGB color image and enlarge the image to twice the size using linear interpolation.   

\subsection{Spectrum Correction}
We observe a different frequency response across devices which results in a performance degradation for under-represented devices. Hence, we calibrate the features of the audio segments by applying spectrum correction instead of training or fine-tuning the model for a specific device~\cite{kochetov2018noise},~\cite{gairola2020respirenet}. The spectrum correction or calibration proposed in~\cite{nguyen2020acoustic}, which was first applied for acoustic scene classification, scales the frequency response of the recording devices. In particular, the calibration coefficients are calculated for each device based on data from reference devices. Table~\ref{device_sample} shows the recorded data portions of each recording device of the ICBHI dataset. The magnitude spectrum $\boldsymbol{s}^{k}_{i}$ of each segment $i$ recorded by the device $k$ is an averaged spectrum along the time axis of all FFT windows. The mean device spectrum $\bar{\boldsymbol{s}}_k = \frac{1}{N_k}\sum_{i=1}^{N_k}{{\boldsymbol{s}}^k_i}$, where device $k$ records $N_k$ segments corresponding to $N_k$ spectra. The reference spectrum $\boldsymbol{s}_{ref}$ is furthermore averaged over all mean device spectra of the $D$ reference devices $\boldsymbol{s}_{ref}=\frac{1}{|D|}\sum_{k \in D}{\bar{\boldsymbol{s}}_k}$, where $D$ contains the indices of the reference devices. We investigate different cases of reference devices based on their prominence i.e only one device either \textit{AKGC417L} or \textit{Meditron} or both \textit{AKGC417L} and \textit{Meditron}, or all recording devices. The scaling coefficients of each device ($\boldsymbol{c}_{k}$) is the element-wise fraction (i.e for each frequency bin) of the reference spectrum and its corresponding device spectrum $\boldsymbol{c}_{k} = \frac{\boldsymbol{s}_{ref}}{\bar{\boldsymbol{s}}_k}$. The magnitude of the STFTs of each device is scaled by using the corresponding coefficient vector $\boldsymbol{c}_{k}$ for the frequency bins. We empirically observed that the normalization in spectrogram domain is more successful than in log-mel domain.

\begin{table}[t]
    \renewcommand{\arraystretch}{1.3}
    \caption{Percentage of samples recorded by each device of the ICBHI dataset}
    \label{device_sample}
    \centering
    
    \begin{tabular}{c c c c c}
        \hline
        \textbf{Device} & \textbf{AKGC417L} & \textbf{Meditron} & \textbf{Litt3200} & \textbf{LittC2SE}\\
        \hline
        \textbf{\% Sample} & 63\% & 21\% & 9\% & 7\% \\
        \hline
    \end{tabular}
\end{table}

\subsection{Data Augmentation}
The ICBHI 2017 dataset is extremely imbalanced with around 53\% of respiratory cycles belonging to the normal class and 86\% of audio recordings belonging to COPD. Furthermore, with our multi-channel lung sound dataset, around 71\% of respiratory cycles are annotated as normal class. Therefore, we use data augmentation in both time domain and time - frequency domain in order to balance the training dataset and prevent over-fitting. 
\subsubsection{Time Domain} 
For ALSC of the ICBHI dataset, we use time stretching to increase/reduce the sampling rate of an audio signal without affecting its pitch~\cite{driedger2016review}. It is used to double the number of segments of the wheeze, and both wheeze and crackle classes. We use a random sampling rate uniformly distributed with $\pm$10\% of the original sampling rate. For RDC of ICBHI, time stretching is used for all classes to double the number of samples. Furthermore, on the doubled training set further data augmentation methods\footnote{https://github.com/makcedward/nlpaug} i.e volume adjusting, noise addition, pitch adjusting and speed adjusting are randomly applied based on a predefined probability.

\subsubsection{Time-Frequency Domain}
\begin{itemize}
	\item Vocal tract length perturbation (VTLP) selects a random wrap factor $\alpha$ for each recording and maps the frequency $f$ of the signal bandwidth to a new frequency $f'$~\cite{jaitly2013vocal}. 
    We select $\alpha$ from a uniform distribution $\alpha \sim \mathcal{U}(0.9, 1.1)$ and set the maximum signal bandwidth to $F_{hi}$ = [3200, 3800]. VTLP is applied directly to the mel filter bank rather than distorting each spectrogram. VTLP is applied to enlarge the dataset for all classes in both tasks for both the original training set and the time stretched data.
	\item Additionally, we double the log-mel features by adding the flipped log-mel features (in frequency axis) for the ALSC and crackle detection task of our dataset.  
\end{itemize}

\subsection{Exploiting Transferred Knowledge}
\subsubsection{Transfer Learning}
Deep neural networks (DNNs) trained from scratch require large amounts of data. As data collecting is a time consuming task for lung sound data, transferring pre-trained parameters from DNNs, which are trained on other datasets e.g. ImageNet is advantageous. Less data of the target task is required, faster training is enabled, and usually better performance after fine-tuning the model on the target task is achieved~\cite{he2019rethinking}. Therefore, fine-tuning brings great benefit to the research community.   

Given a DNN $M_0$ pre-trained on a source dataset $\mathcal{D}_s=\{(x^{i}_{s}, y^{i}_s)\}^{m_s}_{i=1}$, transfer learning aims to fine-tune $M_0$ on a target dataset $\mathcal{D}_t=\{(x^{i}_{t}, y^{i}_t)\}^{m_t}_{i=1}$. In this work, $\mathcal{D}_s$ is selected from ImageNet and $\mathcal{D}_t$ is the ICBHI 2017 dataset or our multi-channel lung sound dataset. Only $\mathcal{D}_t$ and the pre-trained model $M_0$ are available during fine-tuning. 
Because $\mathcal{D}_s$ and $\mathcal{D}_t$ are different domains, which may have different input spaces $\mathcal{X}_s$ and  $\mathcal{X}_t$, corresponding to different output spaces $\mathcal{Y}_s$ and  $\mathcal{Y}_t$, respectively. Therefore, $M_0$ can not be directly applied to the target data. It is common practice, to split $M_0$ into two parts: a general representation function $F_{{\bar{\theta}}^0}$ (parametrized by ${\bar{\theta}}^0$) and a task-specific function $G_{\theta^0_s}$ (parametrized by $\theta^0_s$), which denotes the last layers of the pre-trained model. Usually, the representation function is retained and the task-specific function is replaced by a randomly initialized function $H_{\theta_t}$ (parameterized by $\theta_t$) whose output space matches $\mathcal{Y}_t$. Hence, we optimize 
\begin{equation}\label{tfloss}
    ({\bar{\theta}}^*, \theta^*_t) = \underset{\bar{\theta},\theta_t }{\text{argmin}}\frac{1}{\left | \mathcal{D}_t \right |}\sum_{i=1}^{m_t}l(H_{\theta_t}(F_{\bar{\theta}}(x^i_t)), y^i_t),
\end{equation}
where $l(\cdot)$ is a loss function such as cross-entropy for classification. We will call this vanilla fine-tuning. Pre-trained parameters ${\bar{\theta}}^0$ provide a good starting point for the optimization. It means that the vanilla fine-tuning for a target dataset can be beneficial by transferring the knowledge of the part $F_{{\bar{\theta}}^0}$ of the source dataset.

In this work, we explore different depths of ResNet architectures i.e. ResNet18,  ResNet34,  ResNet50 and  ResNet101 as neural network backbones.

\subsubsection{Co-tuning}
Co-tuning for transfer learning enables full knowledge transfer of the   pre-trained models using a two-step framework~\cite{you2020co}. The first step is learning the relationship between source categories and target categories from the pre-trained model with calibrated predictions. Secondly, target labels (one-hot labels) and source labels (probabilistic labels) translated by the category relationship, collaboratively supervise the fine-tuning process. Co-tuning empirically proves its ability in enhancement of the performance compared to vanilla fine-tuning of the ImageNet pre-trained models~\cite{you2020co}. In this work, we apply co-tuning to fully exploit the ImageNet pre-trained models for significantly distinct datasets such as the ICBHI  and our multi-channel lung sound dataset. The co-tuning block in Fig.~\ref{fig:framework} shows the source output layer $G_{\theta_s}$, the target output layer $H_{\theta_t}$, the ResNet50 backbone $F_{\bar{\theta}}$ and category relationship, which is the relationship between output spaces i.e. the conditional distribution $p(y_s|y_t)$. 

In the training, the category relationship $p(y_s|y_t)$ is needed to translate target labels $y_t$ into probabilistic source categories $y_s$, which is use to fine-tune the  task-specific function $G_{\theta^0_s}$. The gradient of $G_{\theta^0_s}$ can be back-propagated into $F_{\bar{\theta}}$. Both outputs $y_t$ and $y_s$ collaboratively supervise the transfer learning  process described as
\begin{equation}\label{ctloss}
    \begin{split}
        (\bar{\theta}^*, \theta_{t}^{*}, \theta_{s}^{*})=\underset{\bar{\theta},\theta_t, \theta_s }{\text{argmin}}\frac{1}{\left | \mathcal{D}_t \right |}\sum_{i=1}^{m_t} 
        [ l(H_{\theta_t}(F_{\bar{\theta}}(x^i_t)), y^i_t) \\
        + \lambda l(G_{\theta_s}(F_{\bar{\theta}}(x^{i}_{t})),p(y_s|y_t=y_{t}^{i})) ],
    \end{split}
\end{equation}
where $\lambda$ trades off the target and source supervisions. Variables $\bar{\theta}$ and $\theta_s$ are initialized from pre-trained weights $\bar{\theta}^0$ and $\theta^0_s$. In this way, the pre-trained parameters $\bar{\theta}^0$, $\theta^0_s$ are fully exploited in the collaborative training. During inference, the task specific layers $G_{\theta^0_s}$ are removed to avoid the additional cost.

The category relationship $p(y_s|y_t)$ is computed based on the output of task-specific function $G_{\theta^0_s}$ (i.e. a probability distribution over source categories $\mathcal{Y}_s$) and target labels $\mathcal{Y}_t$ by two ways: 
\begin{itemize}
	\item Direct approach: The category relationship is determined as average of the predictions of the pre-trained source model over all samples of each target category i.e.
\begin{equation}\label{directpysyt}
    p(y_s|y_t=y)\approx \frac{1}{|\mathcal{D}_t|}\sum_{ (x, y_t) \in \mathcal{D}_t|y_t=y } {M_0(x)},
\end{equation}
where the pre-trained model $M_0$ is considered as a probabilistic model approximating the conditional distribution $M_0(x)\approx p(y_s|x)$. 

\item Reverse approach: When categories in the pre-trained dataset are diverse enough to compose a target category, we can use a reverse approach. We learn the mapping $y_s \rightarrow  y_t$ from $(M_0(x_t), y_t)$ pairs, where $y_t$ is target labels and $M_0(x)\approx p(y_s|x)$ is a probability distribution over source categories $\mathcal{Y}_s$. Then $p(y_s|y_t)$ can be calculated from $p(y_t|y_s)$ by Bayes's rule.

\end{itemize}

In addition, according to~\cite{you2020co}, it is necessary to calibrate the neural network, i.e. calibrating the probability output of the pre-trained model, to enhance performance.

\subsubsection{Stochastic Normalization (StochNorm)}
In~\cite{kou2020stochastic}, stochastic normalization is proposed to avoid over-fitting during fine-tuning on small dataset. It replaces the Batch Normalization (BN) layers. It implement a two-branch architecture including one branch normalized by mini-batch statistics and another branch normalized by moving statistics (specified in detail below). 
A stochastic selection mechanism like Dropout is used between the two branches to avoid over-depending on some sample statistics. This is interpreted as an architecture regularization. 

Furthermore, StochNorm uses the moving statistics of the pre-trained model for the initial statistics to better exploit prior knowledge in pre-trained networks.

Given a mini-batch of feature maps of each channel $z=\{z_i\}^m_{i=1}$ and a moving statistic update rate $\alpha \in (0,1)$. The normalization process in the two branches is calculated as 
\begin{equation}\label{z0z1}
   \hat{z}_{i, 0}=\frac{z_i-\tilde{\mu}}{\sqrt{\tilde{\sigma}^{2}+\varepsilon }}, \quad    \hat{z}_{i, 1}=\frac{z_i-\mu}{\sqrt{\sigma^{2}+\varepsilon }},
\end{equation}
during training, where the mean $\mu$ and variance $\sigma^2$ of the current mini-batch data of size $m$ 
\begin{equation}\label{musigma}
    \mu \leftarrow  \frac{1}{m}\sum_{i=1}^{m}z_i,  \quad  
    \sigma ^{2} \leftarrow \frac{1}{m}\sum_{i=1}^{m}(z_i -\mu)^{2}
\end{equation}
are used as usual, while the other branch uses moving statistics $\tilde{\mu}$ and $\tilde{\sigma}^2$ of the training data 
\begin{equation}\label{movingmusigma}
    \tilde{\mu} \leftarrow \tilde{\mu}+\alpha(\mu-\tilde{\mu}), \quad  
    \tilde{\sigma}^{2} \leftarrow \tilde{\sigma}^{2}+\alpha(\sigma^{2} - \tilde{\sigma}^{2}).
\end{equation}
The moving statistics are initialized by using corresponding parameters from the pre-trained model. During forward propagation, either $\hat{z}_{i, 0}$ or $\hat{z}_{i, 1}$ is randomly selected with probability $p$ in each channel of the normalization layers and each step of training, i.e. 
\begin{equation}\label{xhat}
    \hat{z}_i = (1-s)\hat{z}_{i, 0} + s\hat{z}_{i, 1},
\end{equation}
where $s$ is the branch-selection variable generated from a Bernoulli distribution $s \sim \textbf{Bernoulli}(p)$. The learnable scale and shift parameters $\beta$, $\gamma$ can be applied after the stochastic selection as usual
\begin{equation}\label{yi}
    y_i \leftarrow \gamma \hat{x}_i + \beta.
\end{equation}
Stochastic normalization in Fig.~\ref{fig:framework} uses a ResNet backbone where BN layers are replaced by StochNorm.

\subsubsection{Combination of Co-tuning and Stochastic Normalization}
 We empirically evaluate the combination of co-tuning and StochNorm for lung sound classification. To do that, the category relationship is initially calculated based on the pre-trained ResNet models, followed by replacing BN layers with the StochNorm modules in the backbone of the original co-tuning architecture (i.e replacing the green ResBlocks from Co-tuning by the blue ResBlocks of Stochastic Normalization in Fig.~\ref{fig:framework}). After that, the fine-tuning of the co-tuning is processed on the new architecture.

\section{Experiments}
\label{sec:experiments}
In this section, we first provide details of the experimental setup. Furthermore, we empirically evaluate the following cases:
\begin{itemize}
    \item Transfer learning of different pre-trained ImageNet ResNet models on the ICBHI dataset.
	\item Ablation study for respiratory segment length, spectrum corrections and flipping data augmentation.
	\item Transfer learning of different ResNet models pre-trained on ImageNet and ICBHI for our multi-channel lung sound dataset.
\end{itemize}
Our systems for the ALSC and RDC tasks on the ICBHI dataset are also compared against state-of-the-art works for the official ICBHI data split and five-fold cross-validation. Additionally, we compare our best system for crackle detection to our previous work on the multi-channel lung sound dataset.

\subsection{Evaluation Metrics}
We use the evaluation metrics supported by the ICBHI Challenge~\cite{rocha2018alpha} for ALSC of 4 classes. The evaluation is based on respiratory cycles using sensitivity (\textit{SE}), specificity (\textit{SP}), average score (\textit{AS}), known as the average of the sensitivity and the specificity, and the harmonic score (\textit{HS}), known as the harmonic mean of the sensitivity and the specificity. For 2 classes, we determine \textit{SE} and \textit{SP} as in~\cite{dPerna2019lstm} and~\cite{pham2021cnn} and \textit{AS} and \textit{HS} as in~\cite{rocha2018alpha}.   

Similarly, for RDC of 3 classes and 2 classes, a recording-wise evaluation is performed using \textit{SE} and \textit{SP} as in~\cite{dPerna2019lstm} and~\cite{pham2021cnn} and \textit{AS} and \textit{HS} as in~\cite{rocha2018alpha}.   

Furthermore, for our multi-channel lung sound dataset, we calculate Precision ($P_+$), Sensitivity or Recall ($Se$), and the F$_1$-score ($F_1$) as specified in~\cite{messner2018crackle}.
Precision  provides information about how many of the respiratory cycles recognized as crackles are actually true. Sensitivity  provides information about how many respiratory cycles containing crackles are actually recognized as crackles. The F$_1$-score is the harmonic mean of precision and sensitivity.

\subsection{Experimental Setup}
We evaluate our ALSC system for 4 and 2 classes on the official ICBHI 2017 dataset split, which consists of 60\% recordings for the training set and 40\% for the test set. Each patients is either in the training or test set. The reported performance is the average accuracy of five independent runs.
For the RDC task of 3 and 2 classes, our proposed system is evaluated on both the official dataset split and five-fold cross-validation. Again, data of each patients is not shared between folds. For five-fold cross-validation, one fold is used as test set, the remaining folds are used for training. As co-tuning requires a validation set to compute the category relationship, we randomly select 20\% of the samples from the training set. For cross-validation, the average of the best performance on the test sets is represented. 

Due to the limited amount of data samples in our multi-channel lung sound dataset, we use 7-fold cross-validation with the recordings of each IPF subject appearing once in the test set. Each subject is assigned to either training, validation or test set. The best model is selected based on the best accuracy on the validation set. The reported performance of the system is an average accuracy of seven folds using the same data splittings. 

Experiments are implemented based on Pytorch~\cite{NEURIPS2019_bdbca288}. 
For vanilla fine-tuning, the learning rate and number of epochs is set to 0.001 and 150 for all tests, respectively. The fine-tuning of co-tuning and stochastic normalization techniques\footnote{Code are available at https://github.com/thuml/Cotuning and https://github.com/thuml/StochNorm.} updates the weights after each mini-batch. The learning rate of the feature representation layers and the last layer are set to 0.001 and 0.01, respectively. The fine-tuning process optimizes the cross entropy loss using SGD with a momentum of 0.9. The batch size is 32 for all experiments.

\subsection{Experimental Results}

\subsubsection{Transfer learning techniques for different ResNets}
\begin{figure}[t]
	\centering
    \includegraphics[width=0.5\textwidth]{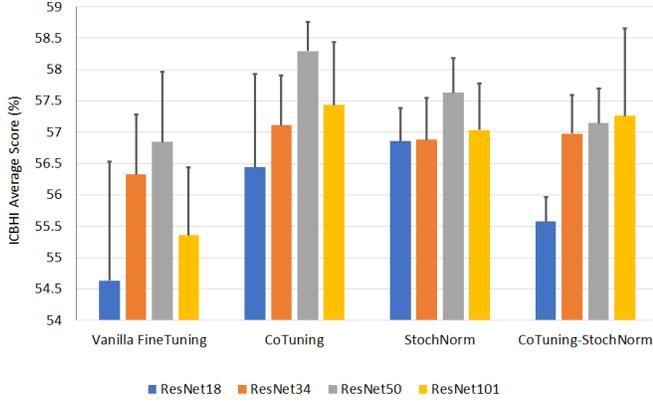}
	\caption{Comparison of vanilla transfer learning, co-tuning, stochastic normalization and both co-tuning and stochastic normalization of different ResNet backbones for the adventitious lung sound classification task of 4 classes.}
	\label{fig:finetuningcomparision}
\end{figure}

\begin{figure}[t]
	\centering
    \includegraphics[width=0.5\textwidth]{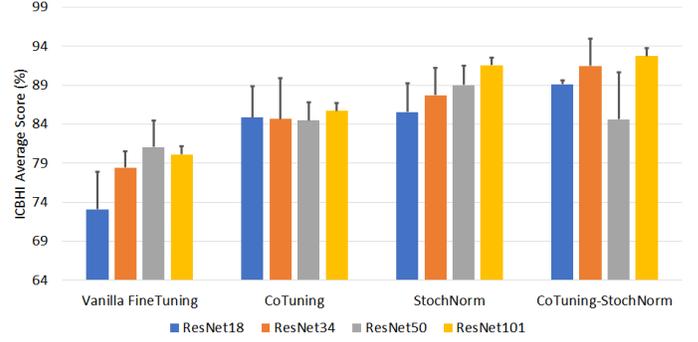}
	\caption{Comparison of vanilla transfer learning, co-tuning, stochastic normalization and both co-tuning and stochastic normalization of different ResNet backbones for the respiratory disease classification task of 3 classes.}
	\label{fig:finetuningcomparision_disease}
\end{figure}

We evaluate the vanilla fine-tuning (\textit{VanillaFineTuning}), co-tuning (\textit{CoTuning}), stochastic normalization (\textit{StochNorm}) and the combination of co-tuning and stochastic normalization (\textit{CoTuning-StochNorm}) for different ResNet architectures trained on the ImageNet dataset for the ALSC task of 4 classes (see Fig.~\ref{fig:finetuningcomparision}) and the RDC task of 3 classes (see Fig.~\ref{fig:finetuningcomparision_disease}) on the official ICBHI dataset split. These systems use a segment length of 8s, spectrum correction using reference data $\boldsymbol{s}_{ref}$ of all devices and all data augmentation methods introduced in Section C. 

Fig.~\ref{fig:finetuningcomparision} shows that ResNet50 is the best performing architecture to build the backbone for these transfer learning techniques of the 4-class ALSC task. ResNet101 is also performing well except for vanilla fine-tuning. Co-tuning achieved the best performance of $\sim$58\% compared to the other techniques. Although \textit{CoTuning} and \textit{StochNorm} improve significantly the performance of \textit{VanillaFineTuning}, the combination of co-tuning and StochNorm is not able to outperform the original techniques for this task.  

\begin{figure}[t]
	\centering
    \includegraphics[width=0.5\textwidth]{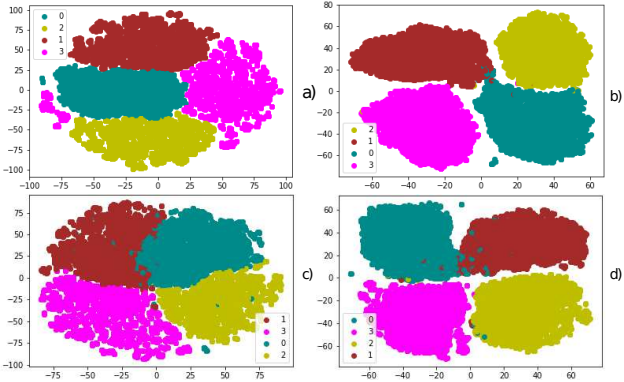}
	\caption{Average pooling output representations reduced into 2D by t-distributed stochastic neighborhood embedding (t-SNE) of the ResNet50 backbone architecture. Distributions of training set of a) vanilla fine-tuning, b) co-tuning, c) stochastic normalization d) combination of co-tuning and stochastic normalization. The color indicates the classes.}
	\label{fig:tsnecomparision}
\end{figure}

Fig.~\ref{fig:tsnecomparision} visualizes of the average pooling outputs of the ResNet50 architecture for different transfer learning techniques projected to 2D by t-distributed stochastic neighbourhood embedding (t-SNE)~\cite{JMLR:v9:vandermaaten08a}. Distributions of the training set using vanilla fine-tuning, co-tuning, stochastic normalization and combination of co-tuning and stochastic normalization are shown at a), b), c) and d), respectively.  Comparing to vanilla fine-tuning (a) and stochastic normalization technique (c), the distribution of 4 classes using co-tuning (b) and the combination of co-tuning and stochastic normalization (d) bring a large margin. It shows that the collaborative fine-tuning using the category relationship of source and target domain is useful for the adventitious lung sound classification task.

In Fig.~\ref{fig:finetuningcomparision_disease}, we see that the different transfer learning techniques using the ResNet101 model achieve the best performance for the 3-class RDC task. The ResNet50 model works better than others for the vanilla fine-tuning. \textit{CoTuning-StochNorm} and \textit{StochNorm} achieved a better performance compared to \textit{CoTuning} and \textit{VanillaFineTuning}. It proves the efficiency of the stochastic normalization in the fine-tuning process for the RDC task.  

\subsubsection{Respiratory segment length}
The length of respiratory cycles in the ICBHI dataset varies in a wide range. Hence, we applied cycle splitting into segments and perform sample padding in order to obtain fixed-length segments. We observe different segment lengths for the ResNet 50 model fine-tuned by co-tuning and applied data augmentation in both time domain and time-frequency domain with spectrum correction. Results are shown in Table 3. The best score is obtained with 8s fixed-length segments (AS) for 4 classes. We also use 8s as the fixed length for other tasks of the ICBHI and our lung sound dataset.

\begin{table}[t]
        \renewcommand{\arraystretch}{1.3}
        \caption{Respiratory segment length: average score (AS) for various input length sizes using co-tuning of ResNet50 as backbone network and data augmentation without spectrum calibration.}
        \label{fixedlength}
        \centering
        \scalebox{0.85} {
        \begin{tabular}{c c c c c c c}
            \hline
            \textbf{Length.}  & \textbf{4 sec} & \textbf{5 sec} & \textbf{6 sec} & \textbf{7 sec} & \textbf{8 sec} & \textbf{9 sec} \\
            \hline
            \textbf{AS} & 54.22 & 56.25 & 56.13 & 56.55 & \textbf{56.67} & 56.58\\
               & $\pm$ 2.27 & $\pm$ 1.71 & $\pm$ 1.17 & $\pm$ 0.91 & $\pm$ \textbf{1.16} & $\pm$ 1.03\\
            \hline
        \end{tabular}}
    \end{table}

\subsubsection{Spectrum correction}
We experiment on the ICBHI dataset without spectrum calibration and with spectrum calibration using different reference spectra $\boldsymbol s_{ref}$, which are determined by one or more devices. \textit{No-Calib} denotes that no spectrum correction is applied. \textit{Calib-Dev1} and \textit{Calib-Dev2} denote calibration using data of device AKGC417L and Meditron, respectively. \textit{Calib-Dev1Dev2} denotes calibration using data of both devices of AKGC417L and Meditron and \textit{Calib-AllDev} denotes spectrum adaptation using reference data of four devices. From Table~\ref{spectrumcorrection}, we can see that co-tuning of the ResNet50 model using reference data of all devices achieves the best performance, it is 1.62\% (absolute) better than without using spectrum calibration. Thus, we apply spectrum calibration for both adventitious lung sound classification and respiratory diseases classification.
\begin{table}[t]
    \renewcommand{\arraystretch}{1.3}
    \caption{Comparison of spectrum correction methods using different reference data. Co-tuning of the ResNet50 with data augmentation is used.}
    \label{spectrumcorrection}
    \centering
    \scalebox{0.8}{
        \begin{tabular}{c c c c c c}
            \hline
            \textbf{}  & \textbf{No-Calib} & \textbf{Calib-Dev1} & \textbf{Calib-Dev2}  & \textbf{Calib-Dev1Dev2} & \textbf{Calib-AllDev}\\
            \hline
            \textbf{AS}   & 56.67    & 57.85   & 57.85   & 57.34   & \textbf{58.29}\\
            & $\pm$ 1.16  &  $\pm$ 0.40 &  $\pm$ 1.00 &  $\pm$ 0.26 &  $\pm$ \textbf{0.46}\\
            \hline
        \end{tabular}
        }
    \end{table}
\subsubsection{Flipping data augmentation}
\begin{figure}[t]
	\centering
    \includegraphics[width=0.5\textwidth]{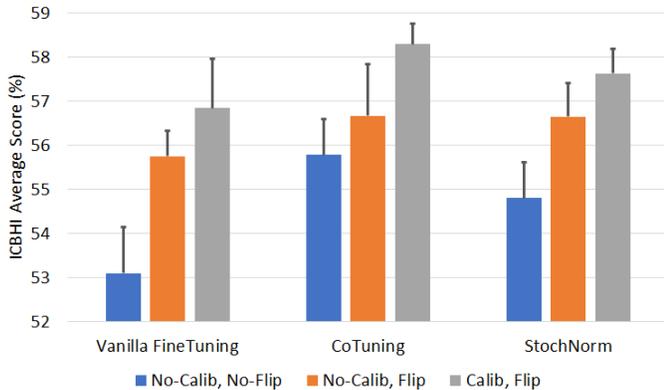}
	\caption{Comparison of vanilla transfer learning, co-tuning and stochastic normalization using ResNet50 for the cases of spectrum correction and flipping data augmentation.}
	\label{fig:calibflipcomparision}
\end{figure}

We apply data augmentation in time domain and VTLP in order to balance the dataset. In this section we focus on the influence of feature flipping data augmentation (see Section III). Fig.~\ref{fig:calibflipcomparision} shows that when the system does not use spectrum correction, doubling the size of the augmented training set by the flipping technique always performs well for vanilla fine-tuning, co-tuning and stochastic normalization. It improves significantly the performance of vanilla fine-tuning and stochastic normalization of about 3\% and 2\%, respectively. For co-tuning, the flipping data augmentation achieves an improvement of 1\% accuracy. Furthermore, we can see from Fig.~\ref{fig:calibflipcomparision} that using the combination of spectrum calibration and flipping data augmentation always enhances the robustness of the adventitious lung sound classification systems. 

\begin{figure}[t]
	\centering
    \includegraphics[width=0.5\textwidth]{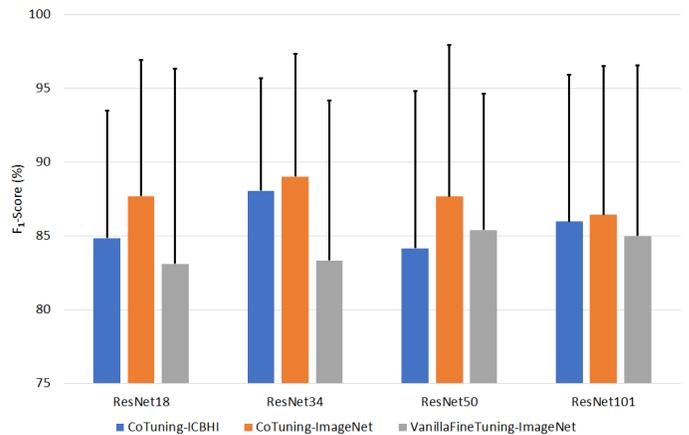}
	\caption{Comparison of co-tuning and vanilla fine-tuning of different ResNet architectures pre-trained on ICBHI and ImageNet for crakle detection. Our multi-channel lung sound dataset and flipping data augmentation are used.}
	\label{fig:elmarcomparision}
\end{figure}

\begin{table}[t]
    \renewcommand{\arraystretch}{1.3}
    \caption{Comparison between the proposed systems and the system in~\cite{nguyen2021crackle} using our multi-channel lung sound dataset for crackle detection task}
    \label{comparisonours}
    \centering
    \scalebox{0.8}{
    \begin{tabular}{l c c c}
        \hline
         \textbf{Method}  & \textbf{Se(\%)}  & \textbf{P+(\%)} & \textbf{F$_1$-Score(\%)}\\
        \hline
        MI-CNN, ICBHI FineTuning~\cite{nguyen2021crackle} & 85.32  & 84.11  & 84.77\\
        \hline
        ResNet34, ICBHI CoTuning (Ours) & 81.42  & 94.70  & 86.81\\
       \textbf{ResNet34, ImageNet CoTuning (Ours)} & \textbf{82.17}  & \textbf{95.88}  & \textbf{87.59}\\
       ResNet50, ImageNet StochNorm (Ours) & 81.82  & 95.22  & 87.42\\
       ResNet50, ImageNet CoTuning-StochNorm (Ours) & 80.84  & 94.90  & 86.46\\
       ResNet50, ImageNet VanillaFineTuning (Ours) & 77.42  & 96.59  & 85.39\\
       \hline
    \end{tabular}
    }
\end{table} 

\subsubsection{Effect of pre-trained model on the multi-channel lung sound dataset}
According to the above evaluation of transfer learning techniques for  different residual neural networks for the 4-class ALSC task, co-tuning achieves the best performance. Thus, we evaluate the effect of pre-trained models using co-tuning (\textit{CoTuning}) for the 2-class ALSC task on our multi-channel lung sound dataset. It is shown in Fig.~\ref{fig:elmarcomparision}. Co-tuning using the ImageNet pre-trained model always outperforms that of the ICBHI pre-trained model. The smaller ResNet architectures tend to work better for co-tuning. We also can see from Fig.~\ref{fig:elmarcomparision} that the ResNet34 backbone system achieves the best performance, followed by ResNet18, ResNet50 and ResNet101. In addition, Fig.~\ref{fig:elmarcomparision}  shows that transferred knowledge from full pre-trained models of ICBHI and the ImageNet dataset by co-tuning to our small lung sound dataset can achieve better accuracy than vanilla fine-tuning (\textit{VanillaFineTuning}) using the ImageNet pre-trained model.\break

Overall, the best segment length for lung sound classification tasks is 8s. Spectrum correction is useful to improve the performance of our ALSC and RDC system on the ICBHI dataset. This helps to correct the different frequency responses of the recording devices. The ALSC system using the flipping data augmentation enhances performance on both ICBHI and our multi-channel lung sound dataset. The new transfer learning methods always outperform vanilla transfer learning. Co-tuning works better for the ALSC task while StochNorm and its combination with the co-tuning achieve higher performance for the RDC task. Furthermore, ResNet34 and ResNet50 are more suitable for the ALSC tasks, while a large ResNet101 model tends to be more robust for the RDC task in most of transfer learning settings.   

\subsection{Performance Comparison}
\subsubsection{Comparison to state-of-the-art systems using the ICBHI dataset}
Table~\ref{comparisonicbhi_alsc} and Table~\ref{comparisonicbhi_rdc} show the comparison of our best systems of different transfer learning techniques and state-of-the-art systems (see Section V for more details on the systems) for the ALSC and RDC tasks, respectively. Our best systems are presented in bold and the highest scores are presented in bold and italic. It is notable that the performances on the official 60/40 ICBHI separation without common patients in both sets are significantly lower than that of randomly 80/20 splitting i.e. 5-fold cross validation and overlap of the same patients in both sets. Despite of the same fixed length for segments, the RDC systems always achieve considerably higher performance compared to the ALSC system for different sub-tasks. The RDC tasks have the full audio recordings which consists of many available respiratory cycles, while the ALSC tasks are processed and evaluated on respiratory cycles. 

We evaluate our proposed system on the official ICBHI split for the  4 and 2 class ALSC tasks. Our best systems of different fine-tuning techniques outperform the other ALSC systems. Our system using co-tuning of the ResNet50 pre-trained model achieve the highest ICBHI average score at 58.29\% and 64.74\% for the 4-class and 2-class ALSC task, respectively. Our RDC systems are evaluated on the official ICBHI split and the 5-fold cross-validation method. On the official dataset split of the 3-class RDC task, our systems achieves the best performance with the ResNet101 pre-trained architecture combined with stochastic normalization. It obtains 92.72\% of the ICBHI average score. While for the 5-fold cross-validation evaluation, the stochastic normalization of the ResNet101 model obtains the best average score 95.73\%. It has an around 5\% better average score compared to the state-of-the-art systems. On the 2-class RDC task, our systems using stochastic normalization achieves the average scores of 93.77\% and 98.20\% for the official splitting and 5-fold cross-validation, respectively. Our best 2-class RDC system outperforms all compared systems on the former evaluation, but it scores 1.02\% lower compared to the system in~\cite{mukherjee2021automatic}.

\subsubsection{Comparison for our multi-channel lung sound dataset}
Table~\ref{comparisonours} compares our best systems using different transfer learning techniques with our previous system using fine-tuning for a multi-input CNN model~\cite{nguyen2021crackle} on the multi-channel lung sound dataset. We can see that our best transfer learning systems outperform the previous system. The co-tuning system using the ResNet34 model pre-trained on ImageNet achieves the best performance, closely followed by the StochNorm system using the pre-train ResNet50 model. The best F$_1$-score is 2.82\% better than for the multi-input fine-tuned system~\cite{nguyen2021crackle}. 

\begin{table*}[t]
    \renewcommand{\arraystretch}{1.15}
    \caption{Comparison between the proposed systems and state-of-the-art systems for ALSC tasks of 4-class and 2-class.}
    \label{comparisonicbhi_alsc}
    \centering
    \scalebox{0.85}{
    \begin{tabular}{l l l c c c c}
        \hline
        \textbf{Task}  & \textbf{Method} & \textbf{Train/Test} & \textbf{SP(\%)}  & \textbf{SE(\%)} & \textbf{AS(\%)} & \textbf{HS(\%)}\\
        \hline
        \textbf{ALSC, 4-class}   & MFCCs, HMM~\cite{jakovljevic2017hidden}  & official 60/40  & -  & -  & 39.56 & -\\
        \textbf{ALSC, 4-class}   & STFT+Wavelet Spectrogram, SVMs~\cite{serbes2017automated}  & official 60/40  & -  & -  & 49.86 & -\\
        \textbf{ALSC, 4-class}   & Low Level Feature, Decition Tree~\cite{chambres2018automatic}  & official 60/40  & \textit{77.80}  & \textit{48.90}  & 49.98 & 49.86\\
        \textbf{ALSC, 4-class}   & STFT+Wavelet Spectrogram, CNN~\cite{minami2019automatic}  & official 60/40  & 81  & 28  & 54 & 42\\
        \textbf{ALSC, 4-class}   & STFT+Wavelet Spectrogram, Bi-ResNet~\cite{ma2019lungbrn}  & official 60/40  & 69.20  & 31.12  & 52.79 & -\\
        
        \textbf{ALSC, 4-class}   & Gamatone Spectrogram, CNN-MoE, DA~\cite{pham2021cnn}  & official 60/40  & 68  & 26  & 47 & - \\
        \textbf{ALSC, 4-class}   & STFT Spectrogram, ResNet-NonLocal, DA~\cite{ma2020lungrn+}  & official 60/40  & 63.20  & 41.32  & 52.26 & -\\
        
        \textbf{ALSC, 4-class}   & STFT Spectrogram, ResNet-SE-SA~\cite{yang2020adventitious}  & official 60/40  & 81.25  & 17.84  & 49.55 & -\\
        \textbf{ALSC, 4-class}   & Logmel Spectrogram, ResNet-FC, DA, BRC, Device Fine-tuning~\cite{gairola2020respirenet}  & official 60/40  & 72.30  & 40.10  & 56.20 & -\\
        
        \textbf{ALSC, 4-class}   & \textbf{Logmel Spectrogram, ResNet50, DA, SC, Vanilla Fine-tuning (Ours)}   & \textbf{official 60/40}  & \textbf{76.33}  & \textbf{37.37}  & \textbf{56.85} & \textbf{50.11} \\
        \textbf{ALSC, 4-class}   & \textbf{Logmel Spectrogram, ResNet50, DA, SC, StochNorm (Ours)}   & \textbf{official 60/40}  & \textbf{78.86}  & \textbf{36.40}  & \textbf{57.63} & \textbf{49.61} \\
        \textbf{ALSC, 4-class}   & \textbf{Logmel Spectrogram, ResNet50, DA, SC, CoTuning (Ours)}   & \textbf{official 60/40}  & \textit{\textbf{79.34}}  & \textit{\textbf{37.24}}  & \textit{\textbf{58.29}} & \textit{\textbf{50.58}}\\
        \textbf{ALSC, 4-class}   & \textbf{Logmel Spectrogram, ResNet101, DA, SC, CoTuning-StochNorm (Ours)}   & \textbf{official 60/40}  & \textbf{78.55}  & \textbf{35.97}  & \textbf{57.26} & \textbf{49.27}\\
        
        \textbf{ALSC, 4-class}   & MFCCs, NMRNN, Device Training~\cite{kochetov2018noise}  &5 folds   & 75  & 62  & 68.5 & -\\
        \textbf{ALSC, 4-class}   & Mel Spectrogram, CNN-RNN.~\cite{acharya2020deep}  & 5 folds  & 84.14  & 48.63   & 66.38 & -\\
        \textbf{ALSC, 4-class}   & STFT Spectrogram, ResNet-NonLocal, DA~\cite{ma2020lungrn+}  & 5 folds  & 64.73  & 63.69  & 64.21 & -\\
        \textbf{ALSC, 4-class}   & Logmel Spectrogram, ResNet-FC, DA, BRC, Device Fine-tuning~\cite{gairola2020respirenet}  & 5 folds  & 83.30  & 53.70  & 68.50 & -\\
        \textbf{ALSC, 4-class}   & MFCCs, CNN~\cite{perna2018convolutional}  & overlap 80/20  & 77  & 45   & 61 & -\\
        \textbf{ALSC, 4-class}   &  MFCCs, LSTM~\cite{dPerna2019lstm} & overlap 80/20  & 85  & 62  & 74  & -\\
        \textbf{ALSC, 4-class}   & Logmel Spectrogram, CNN Snapshot Ensembles, DA~\cite{nguyen2020lung}  & overlap 80/20  & 87.30  & 69.40  & 78.40 & -\\
        \textbf{ALSC, 4-class}   & Gamatone Spectrogram, Ensemble, DA~\cite{pham2020robust}  & overlap 80/20  & 86  & 73  & 80 & - \\
        \textbf{ALSC, 4-class}   & Gamatone Spectrogram, CNN-MoE, DA~\cite{pham2021cnn}  & overlap 80/20  & -  & -  & 78.6 & - \\

        \hline
        
        \textbf{ALSC, 2-class}   & \textbf{Logmel Spectrogram, ResNet50, DA, SC, Vanilla Fine-tuning (Ours)}   & \textbf{official 60/40}  & \textbf{76.33}  & \textbf{52.12}  & \textbf{64.22} & \textbf{61.87} \\
        \textbf{ALSC, 2-class}   & \textbf{Logmel Spectrogram, ResNet50, DA, SC, StochNorm (Ours)}   & \textbf{official 60/40}  & \textbf{78.86}  & \textbf{49.79}  & \textbf{64.32} & \textbf{60.69} \\
        \textbf{ALSC, 2-class}   & \textbf{Logmel Spectrogram, ResNet50, DA, SC, CoTuning (Ours)}   & \textbf{official 60/40}  & \textit{\textbf{79.34}}  & \textit{\textbf{50.14}}  & \textit{\textbf{64.74}} & \textit{\textbf{61.30}}\\
        \textbf{ALSC, 2-class}   & \textbf{Logmel Spectrogram, ResNet101, DA, SC, CoTuning-StochNorm (Ours)}   & \textbf{official 60/40}  & \textbf{78.56}  & \textbf{48.67}  & \textbf{63.61} & \textbf{59.98}\\
        \textbf{ALSC, 2-class}   & CNN~\cite{gairola2020respirenet}  & 5 folds  & 80.90  & 73.10  & 77.0 & -\\
         \textbf{ALSC, 2-class}   & Logmel, ResNet-FC, DA, BRC, Device Fine-tuning~\cite{gairola2020respirenet}  & 5 folds  & 80.90  & 73.10  & 77.0 & -\\
         
        \textbf{ALSC, 2-class}   &  MFCCs, LSTM~\cite{dPerna2019lstm} & overlap 80/20  & -  & -  & 81  & -\\
         \textbf{ALSC, 2-class}   & Logmel Spectrogram, CNN Snapshot Ensembles, DA~\cite{nguyen2020lung}  & overlap 80/20  & 87.30  & 80.10  & 83.70 & -\\
         \textbf{ALSC, 4-class}   & Gamatone Spectrogram, Ensemble, DA~\cite{pham2020robust}  & overlap 80/20  & 86  & 85  & 86 & - \\
         \textbf{ALSC, 2-class}   & Gamatone Spectrogram, CNN-MoE, DA~\cite{pham2021cnn}  & overlap 80/20  & -  & -  & 84.0 & - \\
         \hline
    \end{tabular}
    }
\end{table*}

\begin{table*}[t]
    \renewcommand{\arraystretch}{1.15}
    \caption{Comparison between the proposed systems and state-of-the-art systems for RDC task of 3-class and 2-class.}
    \label{comparisonicbhi_rdc}
    \centering
    \scalebox{0.85}{
    \begin{tabular}{l l l c c c c}
        \hline
        \textbf{Task}  & \textbf{Method} & \textbf{Train/Test} & \textbf{SP(\%)}  & \textbf{SE(\%)} & \textbf{AS(\%)} & \textbf{HS(\%)}\\
        
        \hline
        \textbf{RDC, 3-class}   & Gamatone Spectrogram, CNN-MoE, DA~\cite{pham2021cnn}  & official 60/40  & -  & -  & 84.0 & - \\
        \textbf{RDC, 3-class}   & \textbf{Logmel Spectrogram, ResNet34, DA, SC, Vanilla Fine-tuning (Ours)}   & \textbf{official 60/40}  & \textbf{65.88}  & \textbf{87.47}  & \textbf{76.68} & \textbf{74.48} \\
        \textbf{RDC, 3-class}   & \textbf{Logmel Spectrogram, ResNet101, DA, SC, StochNorm (Ours)}   & \textbf{official 60/40}  & \textbf{90.59}  & \textbf{92.53}  & \textbf{91.56} & \textbf{91.35} \\
        \textbf{RDC, 3-class}   & \textbf{Logmel Spectrogram, ResNet101, DA, SC, CoTuning (Ours)}   & \textbf{official 60/40}  & \textbf{81.18}  & \textbf{90.22}  & \textbf{85.70} & \textbf{85.23}\\
        \textbf{RDC, 3-class}   & \textbf{Logmel Spectrogram, ResNet101, DA, SC, CoTuning-StochNorm (Ours)}   & \textbf{official 60/40}  & \textit{\textbf{91.77}}  & \textit{\textbf{93.68}}  & \textit{\textbf{92.72}} & \textit{\textbf{92.57}}\\
        
        \textbf{RDC, 3-class}   & MFCCs, CNN~\cite{perna2018convolutional}  & overlap 80/20  & 76  & 89   & 83 & -\\
        \textbf{RDC, 3-class}  &  MFCCs, LSTM~\cite{dPerna2019lstm} & overlap 80/20  & 82  & 98  & 90  & -\\
        \textbf{RDC, 3-class}   & Gamatone Spectrogram, CNN-MoE, DA~\cite{pham2020robust}  & overlap 80/20  & 83  & 96  & 90 & - \\
        \textbf{RDC, 3-class}   & Gamatone Spectrogram, CNN-MoE, DA~\cite{pham2021cnn}  & 5 folds  & 86  & 95  & 91 & - \\
        \textbf{RDC, 3-class}   & \textbf{Logmel Spectrogram, ResNet50, DA, SC, Vanilla Fine-tuning (Ours)}   & \textbf{5 folds}  & \textbf{83.02}  & \textbf{90.40}  & \textbf{86.71} & \textbf{84.85} \\
        \textbf{RDC, 3-class} & \textbf{Logmel Spectrogram, ResNet101, DA, SC, StochNorm (Ours)}   & \textbf{5 folds}  & \textit{\textbf{100}}  & \textit{\textbf{91.47}}  & \textit{\textbf{95.73}} & \textit{\textbf{95.48}} \\
        \textbf{RDC, 3-class}   & \textbf{Logmel Spectrogram, ResNet101, DA, SC, CoTuning (Ours)}   & \textbf{5 folds}  & \textbf{96.67}  & \textbf{90.18}  & \textbf{93.42} & \textbf{9.06}\\
        \textbf{RDC, 3-class}   & \textbf{Logmel Spectrogram, ResNet101, DA, SC, CoTuning-StochNorm (Ours)}  & \textbf{5 folds}  & \textbf{97.78}  & \textbf{91.44}  & \textbf{94.61} & \textbf{94.46} \\
        
        \hline
        
        \textbf{RDC, 2-class}   & Low Level Feature, Decition Tree~\cite{chambres2018automatic}  & official 60/40  & -  & -  & 85 & -\\
        \textbf{RDC, 2-class}   & Gamatone Spectrogram, CNN-MoE, DA~\cite{pham2021cnn}  & official 60/40  & -  & -  & 84.1 & - \\
        \textbf{RDC, 2-class}   & \textbf{Logmel Spectrogram, ResNet34, DA, SC, Vanilla Fine-tuning (Ours)}   & \textbf{official 60/40}  & \textbf{65.88}  & \textbf{96.43}  & \textbf{81.16} & \textbf{77.58} \\
        \textbf{RDC, 2-class}   & \textbf{Logmel Spectrogram, ResNet101, DA, SC, StochNorm (Ours)}   & \textbf{official 60/40}  & \textbf{90.59}  & \textbf{93.90}  & \textbf{92.25} & \textbf{90.02} \\
        \textbf{RDC, 2-class}   & \textbf{Logmel Spectrogram, ResNet101, DA, SC, CoTuning (Ours)}   & \textbf{official 60/40}  & \textbf{81.18}  & \textbf{94.29}  & \textbf{87.73} & \textbf{87.12}\\
        \textbf{RDC, 2-class}   & \textbf{Logmel Spectrogram, ResNet101, DA, SC, CoTuning-StochNorm (Ours)}   & \textbf{official 60/40}  & \textit{\textbf{91.77}}  & \textit{\textbf{95.77}} & \textit{\textbf{93.77}} & \textit{\textbf{93.60}} \\
        \textbf{RDC, 2-class}   & MFCCs, LSTM~\cite{dPerna2019lstm}  & overlap 80/20  & 82  & 99   & 91 & -\\
        \textbf{RDC, 2-class}   & Gamatone Spectrogram, CNN-MoE, DA~\cite{pham2020robust}  & overlap 80/20  & 83  & 99  & 91 & - \\
        \textbf{RDC, 2-class}   & Gamatone Spectrogram, CNN-MoE, DA~\cite{pham2021cnn}  & 5 folds  & 86  & 98  & 92 & - \\
        \textbf{RDC, 2-class}   & LPC, MLP classifier~\cite{mukherjee2021automatic}  & 5 folds  & \textbf{-}  & \textbf{-}  & \textit{\textbf{99.22}} & \textbf{-} \\
        \textbf{RDC, 2-class}   & \textbf{Logmel Spectrogram, ResNet50, DA, SC, Vanilla Fine-tuning (Ours)}   & \textbf{5 folds}  & \textbf{83.02}  & \textbf{96.03}  & \textbf{89.52} & \textbf{87.23} \\
        \textbf{RDC, 2-class}   & \textbf{Logmel Spectrogram, ResNet101, DA, SC, StochNorm (Ours)}   & \textbf{5 folds}  & \textbf{100}  & \textbf{96.41}  & \textbf{98.20} & \textbf{98.15} \\
        \textbf{RDC, 2-class}   & \textbf{Logmel Spectrogram, ResNet101, DA, SC, CoTuning (Ours)}   & \textbf{5 folds}  & \textbf{96.67}  & \textbf{94.46}  & \textbf{95.56} & \textbf{95.28}\\
        \textbf{RDC, 2-class}   & \textbf{Logmel Spectrogram, ResNet101, DA, SC, CoTuning-StochNorm (Ours)} & \textbf{5 folds}  & \textbf{97.78}  & \textbf{95.15}  & \textbf{96.46} & \textbf{96.42}\\
        \textbf{RDC, 2-class}   & MFCCs, DWT, time domain features, RUSBoost-DT~\cite{kok2019novel}  & 50/50  & 93  & 86  & 87.10 & - \\
       
        \hline
    \end{tabular}
    }
\end{table*}
\section{Related Works}
We review recent works on ALSC and RDC using the ICBHI 2017 dataset and binary ALSC works (i.e. crackle detection) using the multi-channel lung sound database. In general, it is difficult to compare the score of some proposed methods for ICBHI as a substantial work does not use the official data splitting or use a different evaluation metric.
\subsection{Lung Sound Classification on ICBHI dataset}
There are two main directions: (i) conventional classifiers for low-level features of time or frequency domain, (ii) deep neural networks and robust machine learning techniques for spectral features. 
\subsubsection{Conventional approach}
Jakovljevic et al.~\cite{jakovljevic2017hidden} used hidden Markov models and Gaussian mixture models for MFCCs, which were applied after spectrum subtraction for noise suppression. Their system achieved an average score of 39.56\% on the official ICBHI train-set split and 49.5\% on 10-fold cross-validation for the ALSC task with 4 classes. 
Serbes et al.~\cite{serbes2017automated} proposed a 4-class ALSC system using support vector machines (SVMs) for STFT and wavelet features. It achieved 49.86\% of average score on the official ICBHI data split. 
Furthermore, Chambres et al.~\cite{chambres2018automatic} proposed a system using a boosted tree method for low-level features of spectral information i.e. bark-bands, energy-bands, mel-bands, MFCCs and other features i.e. rythm features and tonal features for both of the ALSC and binary RDC task. Their performance on the official ICBHI split were 49.63\% and 85\% for the 4-class ALSC and 2-class RDC tasks, respectively. 

A binary RDC system using the RUSBoost algorithm, which combines random under sampling and boosting techniques (i.e decision tree as a base classifier) was also introduced~\cite{kok2019novel}. The input of the classifier are features selected from MFCCs, discrete wavelet transform (DWT) and time domain features. The proposed system was evaluated on their own ICBHI dataset split and achieved 87.1\% of average score. 
In addition, Mukherjee et al.~\cite{mukherjee2021automatic} developed a method to detect patients with respiratory infections. They extracted features based on Linear Predictive Coefficient (LPC) for a multilayer perceptron classifier (MPC). The method was evaluated on the ICBHI dataset using 5-fold cross-validation and achieved 99.22\% of accuracy for the 2-class ALSC task.

\subsubsection{Deep learning approach} 
Deep learning systems use CNNs, Recurrent neural networks (RNNs) and hybrid architectures. They are combined with machine learning techniques such as data augmentation, ensemble methods and transfer learning to enhance robustness. 
For the RNN-based systems, Kochetov et al.~\cite{kochetov2018noise} proposed a system using a noise making RNN (NMRNN) and MFCC features to classify cycles of lung sounds into four categories. The performance was evaluated based on 5-fold cross-validation. It is the first work which considers the effect of the recording devices on the performance. They achieved a score of 64.8\% and 68.5\% with training data from all devices and the most often occurring recording device (i.e. AKGC417L), respectively.
Furthermore, a CNN network for MFCCs was used, which achieved 61\% and 83\% of average score for the 4-class ALSC and 3-class RDC task with random train-test split of 80\% and 20\%~\cite{perna2018convolutional}. In~\cite{dPerna2019lstm}, Perna et. al. also introduced different architectures of RNNs such as long short time memory (LSTM), gated recurrent units (GRU), bidirectional-LSTM (BiLSTM) and bidirectional-GRU (BiGRU) for MFCC features to perform 4-class and 2-class ALSC and the ternary and 2-class RDC task. The results on random train-test ICBHI split of 80\% and 20\% with over-lapping of the same patients in both sets are 74\% and 81\% of average score for the 4-class and 2-class ALSC task, respectively. The average performance of the RDC tasks of 3 and 2 classes is 84\% and 91\%, respectively. 

Furthermore, CNNs or hybrid architectures have been used.
In~\cite{nguyen2020lung}, we proposed a lung sound classification using a snapshot ensemble of CNNs for log-mel spectrograms. We applied temporal stretching and vocal tract length perturbation (VTLP) for data augmentation to deal with the class-imbalance of the ICBHI dataset. Our system achieved 78.4\% and 83.7\% of average score on the random train-test split of 80\% and 20\% with common patients in both sets for the ALSC task of 4 classes and 2 classes, respectively.
Acharya et al.~\cite{acharya2020deep} introduced a deep CNN-RNN model for mel spectrograms to classify adventitious lung sounds into four classes. The performance for 5-fold cross validation evaluation was 66.31\%. When this system was combined with a patient specific model tuning strategy, its performance increased up to 71.81\% of average score. 
Similarly, Pham et al.~\cite{pham2020robust},~\cite{pham2021cnn} introduced lung sound classification systems for anomaly sounds and respiratory diseases. In the first work~\cite{pham2020robust}, they proposed various deep learning architectures mainly based on CNNs and RNNs using gammatone filtered spectrograms. They use a 80\%-20\% dataset split, where data from one subject may exist in both training and test set. An average ensemble of these systems achieved 80\% and 86\% of the average score for the 4-class and 2-class ALSC, respectively. The proposed CNN - mixture of expert (MoE) model was suitable for the RDC task of 3 classes and 2 classes with a performance of 90\% and 91\%, respectively.
In~\cite{pham2021cnn} they proposed a CNN-MoE neural network for different feature types i.e. MFCCs, log-mel, gammatone filter and constant Q transform spectrogram. The gammatone filter spectrogram was suggested for the ALSC tasks, while log-mel spectrogram worked better for the RDC task. The average score of the 4-class ALSC task was 47\% for the ICBHI official dataset split. For 5-fold cross-validation with data of the same patient in both sets, their performance was 78.6\% and 84\% for the ALSC task of 4 classes and 2 classes, respectively. On the 3-class RDC task they achieved 85\% of average score on the ICBHI official dataset split and 91\% on 5-fold cross-validation.

Recently, CNN-based systems from diverse architectures i.e. VGGNets, ResNets have been more and more introduced.
Minami et al.~\cite{minami2019automatic} proposed a 4-class ALSC system using a VGG16 neural network for the combination of STFT spectrogram and scalogram. The performance was 54\% of average score on the official ICBHI dataset. 
Ma et al. proposed two ALSC systems for four classes~\cite{ma2019lungbrn},~\cite{ma2020lungrn+}. The first one used an improved Bi-ResNet deep learning architecture based on STFT and wavelet features. Another system used non-local block ResNet with mixup data augmentation for STFT spectrograms. The proposed systems achieved 50.16\% and 52.26\% on the official data split, respectively. The latter work~\cite{ma2020lungrn+} was also evaluated using 5-fold cross-validation and achieved an average score of 64.21\%.

Yang et al.~\cite{yang2020adventitious} proposed a 4-class ALSC system combining the ResNet18 architecture with Squeeze-and-Excitation and spatial attention blocks using STFT spectrogram features. They obtained 49.55\% of average score on the official ICBHI datset split. 
Demir et al. proposed a 4-class ALSC system using pre-trained models for STFT spectrograms converted into color images. In the first approach~\cite{demir2020convolutional}, the pre-trained model was used as feature extractor and combined to an SVM classifier. In the second approach~\cite{demir2020convolutional}, the pre-trained model was fine-tuned on the ICBHI dataset. They achieved 65.5\% and 63.09\% of accuracy for 10-fold cross-validation, respectively. In~\cite{demir2020classification}, they introduced a parallel pooling CNN model for deep feature extraction. It is combined to a linear discriminant analysis (LDA) classifier and random subspace ensembles (RSE). The performance of the proposed system was 71.5\% for 10-fold cross-valuation. However, the evaluation metrics are different. 

Additionally, Gairola et. al.~\cite{gairola2020respirenet} proposed a RespireNet model based on ResNet34 and fully connected layers with a set of techniques i.e. device specific fine-tuning, concatenation-based augmentation, blank region clipping and smart padding to improve the accuracy. The average score for the 4-class ALSC task was 56.2\% and 68.5\% for the official ICBHI dataset split and 5-fold cross-validation, respectively. They also evaluated the proposed system for the ALSC task of two classes and obtained 77.0\% accuracy on 5-fold cross-validation. 

\subsection{Lung Sound Classification on our multi-channel dataset}
In~\cite{messner2018crackle}, Messner et al. introduced an event detection approach with bidirectional gated recurrent neural networks (Bi-GRNNs) using MFCCs to identify crackles in respiratory cycles. The proposed system was evaluated on the first version of the multi-channel lung sound dataset including  10 lung-healthy subjects and 5 patients with IPF. The performance was 72\% of F-score on 5-fold cross-validation evaluation.

In~\cite{messner2020multi}, a classification framework using lung sound signals of all recording channels was introduced to identify healthy and pathological breathing cycles. Lung sounds of one breath cycle of all recording channels were first transformed into STFT spectrograms. Then, the spectrogram were stacked into one compact feature vector. These features were fed into a CNN-RNN model for classification. Its score was 92\% for 7-fold cross-validation evaluation.

We proposed a multi-input CNN model based on transfer learning for the ALSC task of crackles and normal sounds, namely crackle detection~\cite{nguyen2021crackle}  on the multi-channel lung sound classification dataset. The multi-input CNN model shares the same network architecture of the pre-trained CNN model trained on the ICBHI dataset for respiratory cycles and their corresponding respiratory phases. Our system achieved an F-score of 84.71\% on 7-fold cross-validation evaluation.

\section{Conclusion}
We propose robust fine-tuning frameworks to classify adventitious lung sounds and recognize respiratory diseases from lung auscultation recordings using the ICBHI and our multi-channel lung sound datasets. Transferred knowledge of pre-trained models from different ResNet architectures are exploited by vanilla fine-tuning, co-tuning, stochastic normalization and the combination of co-tuning and stochastic normalization techniques. Furthermore, spectrum correction and flipping data augmentation are introduced to improve the robustness of our system. Empirically, our proposed systems outperform almost all state-of-the-art systems for adventitious lung sound and respiratory disease classification. In addition, we also evaluate our adventitious lung sound classification approach using co-tuning on our multi-channel lung sound dataset to detect crackles using different pre-trained models of the ImageNet and ICBHI dataset. The best co-tuning system for 2-class lung sound classification achieves a better performance (2.82\%) compared to our previous work using a multi-input convolutional neural network. We also review state-of-the-art classification systems for adventitious lung sounds and respiratory diseases using the ICBHI dataset and our multi-channel lung sound dataset.


%

\section*{Acknowledgment}
This research was supported by the Vietnamese - Austrian Government scholarship and by the Austrian Science Fund (FWF) under the 
project number I2706-N31. We acknowledge NVIDIA for providing GPU computing resources. The authors would like to thank our colleague, Alexander Fuchs for feedback and fruitful discussions. 

\ifCLASSOPTIONcaptionsoff
  \newpage
\fi



%

\bibliographystyle{IEEEbib}
\bibliography{refs}
%

\end{document}